# Optimal Sizing of On-site Renewable Resources for Offshore Microgrids

Ann Mary Toms, *Student Member, IEEE*, Xingpeng Li, *Senior Member, IEEE*, and Kaushik Rajashekara, *Fellow, IEEE*

*Abstract*—The offshore oil and natural gas platforms, mostly powered by diesel or gas generators, consume approximately 16TWh of electricity worldwide per year, which emits large amount of $CO_2$. To limit their contribution to climate change, a proposed solution is to replace the traditional fossil fuel based energy resources with offshore clean energy. One of the main challenges in designing such a system is to ensure that energy demand is met while minimizing cost and reducing environmental impact. To address this challenge, several strategies including microgrid systems consisting of offshore wind turbines, wave energy converters, tidal energy converters, floating photovoltaic systems and battery energy storage systems are being proposed. In this paper, cost optimization for sizing these renewable energy sources is investigated. A cost optimization renewable sizing (CORS) model is proposed to optimize the sizes of the generation and storage resources. The proposed CORS model considers the variability of the power outputs of various renewable energy sources and load, as well as the cost of different generation technologies and the energy storage system. Simulations conducted on three test systems show the proposed resource sizing method significantly reduces the total lifetime cost of energy while maintaining a high level of reliability and sustainability.

## NOMENCLATURE

**Sets**
$T$ — Set of Time Intervals.

**Indices**
$t$ — Time Intervals.

**Parameters**
$C_{BESS}^{capital}$ — Capital cost of each kWh of BESS.
$C_{FPV}^{capital}$ — Capital cost of an FPV unit.
$C_{OWT}^{capital}$ — Capital cost of an OWT unit.
$C_{TEC}^{capital}$ — Capital cost of a TEC unit.
$C_{WEC}^{capital}$ — Capital cost of a WEC unit.
$C_{BESS}^{decom}$ — Decommissioning cost for each kWh of BESS.
$C_{FPV}^{decom}$ — Decommissioning cost of an FPV unit.
$C_{OWT}^{decom}$ — Decommissioning cost of an OWT unit.

$C_{TEC}^{decom}$ — Decommissioning cost of a TEC unit.
$C_{WEC}^{decom}$ — Decommissioning cost of a WEC unit.
$C_{BESS}^{O\&M}$ — O&M cost of each kWh of the BESS unit per annum.
$C_{FPV}^{O\&M}$ — O&M cost associated with each FPV unit per annum.
$C_{OWT}^{O\&M}$ — O&M cost associated with each OWT unit per annum.
$C_{TEC}^{O\&M}$ — O&M Cost associated with each TEC unit per annum.
$C_{WEC}^{O\&M}$ — O&M cost associated with each WEC unit per annum.
$C_{BESS}^{precom}$ — Pre-commissioning cost of each kWh of BESS.
$C_{FPV}^{precom}$ — Pre-commissioning cost of each FPV Unit.
$C_{OWT}^{precom}$ — Pre-commissioning cost of each OWT Unit.
$C_{TEC}^{precom}$ — Pre-commissioning cost of each TEC Unit.
$C_{WEC}^{precom}$ — Pre-commissioning cost of each WEC Unit.
$P_{Max}^{Char}$ — Maximum Charging Power of BESS.
$P_{Max}^{Disc}$ — Maximum Discharging Power of BESS.
$P_t^{FPV}$ — Available solar power at time period $t$.
$P_t^{Load}$ — Load at time period $t$.
$P_t^{OWT}$ — Available wind power at time period $t$.
$P_t^{TEC}$ — Available tidal power at time period $t$.
$P_t^{WEC}$ — Available wave power at time period $t$.
$SOC_{min}$ — Minimum state of charge of BESS.
$SOC_{max}$ — Maximum state of charge of BESS.
$T_e$ — Expected lifetime of the OHRES system.
$\delta_{BESS}^{degrad}$ — Battery Degradation Cost Factor.
$\eta_{BESS}^{Char}$ — Charging Efficiency of the BESS.
$\eta_{BESS}^{Disc}$ — Discharging Efficiency of the BESS.

**Variables**
$E_{BESS}$ — Total energy capacity of the BESS unit.
$E_{BESS}^{initial}$ — Initial energy capacity of the BESS unit.
$E_{BESS}^t$ — Energy capacity of the BESS unit at time period $t$.
$N_{FPV}$ — Number of floating photovoltaic panels.
$N_{OWT}$ — Number of offshore wind turbines.
$N_{TEC}$ — Number of tidal energy converters.
$N_{WEC}$ — Number of wave energy converters.
$P_t^{Char}$ — Charging power of BESS at time $t$.
$P_t^{Curt}$ — Renewable power curtailment at time period $t$.
$P_t^{Disc}$ — Discharging power of BESS at time period $t$.
$U_t^{Char}$ — Charging status of BESS at time period $t$. 1 represents charging mode and 0 represents discharging or idle mode.
$U_t^{Disc}$ — Discharging status of BESS at time period $t$. 1 represents charging mode and 0 represents discharging or idle mode.

## I. INTRODUCTION

OFFSHORE oil and gas (O&G) production plays a pivotal role in the energy landscape of many countries. However, it is worth noting that these operations are quite energy intensive, with some larger and more complex platforms requiring several hundreds of megawatts of power to keep the

Ann Mary Toms, Xingpeng Li and Kaushik Rajashekara are with the Department of Electrical and Computer Engineering at the University of Houston, Houston, TX 77004, USA (e-mails: atoms2@cougarnet.uh.edu; xli83@Central.UH.EDU, ksraja@Central.UH.EDU).

system running smoothly. Collectively, offshore O&G platforms consume approximately 16TWh of energy per year [1]. In 2019, gas accounted for 21% of global $CO_2$ emissions from fuel, while oil accounted for 34%, with a significant portion coming from offshore O&G rigs [2]. Offshore O&G rigs in Louisiana, Texas, California, and Alaska account for a significant amount of the USA's oil and gas supply.

According to the US Energy Information Administration, the estimated theoretical annual energy that can be extracted from waves of the coasts of the United States is anticipated to be as much as 2640 TWh [3]. According to the National Renewable Energy Laboratory (NREL), the estimated theoretical annual recoverable energy from tidal resources is estimated to be approximately 252 TWh [4]. The potential for offshore wind generation in the US is vast, with more than 2TW estimated [5]. Offshore wind turbines (OWT) are significantly larger than onshore wind turbines. Because offshore winds are stronger and more stable than onshore winds, the capacity factor for offshore wind farms are typically higher [6]-[9]. The US has set a national offshore wind target of 30 GW by 2030 [10]. It is also worth noting that the floating wind farms capable of being built in deep and ultra-deep waters are rapidly developing. Currently, there exists O&G rigs that are already being powered by OWT. The Beatrice wind farm serves as a prime example, featuring two 5MW wind turbines connected to the Beatrice oil production platform in Scotland via submarine cables [11]-[12].

In this paper, an offshore hybrid renewable energy source (OHRES)-based microgrid that integrates wave energy converters (WEC), tidal energy converters (TEC), offshore wind turbines, floating photovoltaic system (FPV) and battery energy storage systems (BESS) to envisage an offshore O&G platform powered solely by renewable sources of energy is considered. A cost optimization for renewable sizing (CORS) model is proposed to optimize the sizes of the generation and storage resources considering various practical factors.

The rest of this paper is organized as follows. Literature review is conducted in Section II. The proposed CORS model for OHRES microgrids is explained in Section III. The dataset is described in Section IV, while the calculations are detailed in Section V. Three test cases and the associated simulation results are presented in Section VI. Section VII concludes the paper. Potential future works are discussed in Section VIII.

## II. LITERATURE REVIEW

Currently, there is an increase in research focused on optimizing the use of renewable energy systems for offshore O&G rigs. Various approaches were studied to meet the electrical demand of O&G rigs by utilizing marine renewable resources, wind energy and solar energy.

A feasibility study assessed the integration of WEC and solar energy system for supplying power to offshore O&G rigs [13]. The results demonstrated that combining these renewable resources led to an increase in electricity production, reduced intra-annual variability, and mitigated intermittency issues. The levelized cost of energy ranged from 140-282 $/MWh

[13]. Similar studies were also performed for O&G rigs in the Caspian sea integrating OWT along with FPV systems [14], and FPV along with ocean thermal energy conversion systems [15]. The results for these studies were also comparable.

Researchers are also interested in the introduction of OWT for meeting the needs of O&G rigs. Preliminary studies were conducted to explore the possibility of replacing fossil fuel powered offshore O&G rigs with OWT in Brazil [16]. Studies were conducted to investigate the possibility of operating four 5 MW OWT in conjunction with gas turbines and BESS [17]-[18]. Additionally, another research proposed the implementation of a 40MW wind farm to power an isolated offshore O&G rig [19]. The simulation results demonstrate that using OWT to power O&G rigs resulted in cost saving by reducing fuel consumption.

Since FPV is a relatively nascent technology, it has greater untapped potential [20]-[23]. Thus, a lucrative alternative is the integration of FPV with other renewables especially OWT [24]. This type of system hybridization enhances the overall productivity of the energy generation system [25]-[27]. Another potential solution is the combination of TEC with WEC and OWT to generate electricity [28]-[29].

One of the research gaps identified is the lack of studies on the optimal sizing of offshore microgrids that incorporate renewable resources such as WEC, TEC, FPV or OWT. It is suggested that future research should focus on developing advanced optimization techniques for optimal design and sizing of microgrids, which would help maximize the use of renewable resources and reduce the reliance on fossil fuels.

Another gap was the need for more comprehensive analysis on the impact of environmental factors such as weather, sea patterns, and energy consumption profiles on the performance of the offshore grids. There is also a need for sensitivity analysis to test the robustness of these microgrids under different scenarios and parameter settings.

## III. CORS MODEL FOR SIZING OHRES MICROGRID

The proposed CORS sizing model for OHRES microgrid is designed to be efficient, flexible, and cost-effective. By using renewable energy sources and smart energy storage solutions, it aims to reduce the carbon footprint of offshore platforms, contributing to a sustainable future.

The proposed CORS model for sizing candidate resources in the OHRES microgrid is composed of (1)-(13), which minimizes the total cost of supplying power for an offshore platform over its lifetime.

The system achieves this objective by considering the costs of five subsystems: WEC, TEC, OWT, FPV and BESS. The ultimate goal is to reduce the total cost of energy while ensuring reliable power supply. The objective function is defined in (1).

$$min\, F(Cost) = f(WEC) + f(TEC) + f(OWT) \\ + f(FPV) + f(BESS) \quad (1)$$

To achieve the goal of minimizing the total cost of maintaining the power supply for offshore platforms, the CORS model considers the cost of each subsystem, including the pre-commissioning cost, capital cost, operations and maintenance costs, decommissioning cost and expected lifetime of the subsystem, which are presented in (2)-(6) for WEC, TEC, OWT, FPV and BESS respectively.

$$f(WEC) = N_{WEC}(C_{WEC}^{precom} + C_{WEC}^{capital} + \{C_{WEC}^{O\&M} \times T_e\} + C_{WEC}^{decom}) \tag{2}$$

$$f(TEC) = N_{TEC}(C_{TEC}^{precom} + C_{TEC}^{capital} + \{C_{TEC}^{O\&M} \times T_e\} + C_{TEC}^{decom}) \tag{3}$$

$$f(OWT) = N_{WT}(C_{OWT}^{precom} + C_{OWT}^{capital} + \{C_{OWT}^{O\&M} \times T_e\} + C_{OWT}^{decom}) \tag{4}$$

$$f(FPV) = N_{FPV}(C_{FPV}^{precom} + C_{FPV}^{capital} + \{C_{FPV}^{O\&M} \times T_e\} + C_{FPV}^{decom}) \tag{5}$$

$$f(BESS) = E_{BESS}(C_{BESS}^{precom} + C_{BESS}^{capital}\{1 + (\delta_{BESS}^{degrad} \times T_e)\} + \{C_{BESS}^{O\&M} \times T_e\} + C_{BESS}^{decom}) \tag{6}$$

Equation (7) presents the power balance equation that takes into account the renewable energy sources, BESS and electrical demand. It ensures a balanced supply and demand of power, optimizing the use of renewable energy while meeting the energy demand of the offshore platforms.

$$(N_{WEC} \times P_t^{WEC}) + (N_{TEC} \times P_t^{TEC}) + (N_{OWT} \times P_t^{OWT}) + (N_{FPV} \times P_t^{FPV}) + (P_t^{Disc} - P_t^{char}) - P_t^{Curt} = P_t^{Load}, \forall t \tag{7}$$

Equations (8)-(13) model the BESS in the OHRES system. These equations enforce constraints to ensure optimal use of the BESS and prevent any energy shortage or wastage. Equation (8) calculates the stored energy of the BESS for each time interval, while (9) ensure the initial and ending energy levels are maintained. Constraint (10) limits the BESS energy level to stay within the maximum and minimum limits, while (11) ensures the BESS cannot charge and discharge in the same time interval. Constraints (12)-(13) regulate the charging and discharging power, preventing it from exceeding the BESS limit.

$$E_{BESS}^t - E_{BESS}^{t-1} = (P_t^{Char} \times \eta_{BESS}^{Char}) - \left(\frac{P_t^{Disc}}{\eta_{BESS}^{Disc}}\right), \forall t \tag{8}$$

$$E_{BESS}^{Initial} = E_{BESS}^{24}, \forall t \tag{9}$$

$$SOC_{min}E_{BESS} \leq E_{BESS}^t \leq SOC_{max}E_{BESS}, \forall t \tag{10}$$

$$U_t^{Char} + U_t^{Disc} \leq 1, \forall t \tag{11}$$

$$0 \leq P_t^{Disc} \leq U_t^{Disc} \times P_{Max}^{Disc}, \forall t \tag{12}$$

$$E_{BESS}^{Initial} = E_{BESS}^{24}, \forall t \tag{13}$$

## IV. Dataset Description

The datasets used for calculating the projected power generation profiles of the four types of renewable resources in the OHRES system are explained below respectively.

### A. Offshore Wind and Wave Data

The dataset used for calculating the output power of the OWT and WEC is obtained from the National Data Buoy Center (NDBC) [30]. This dataset is owned and maintained by NDBC. The datasets have measurements taken at hourly intervals over the course of a year for different locations across USA. Each row in the dataset represents the attributes of meteorological data corresponding to a given time instant. The dataset has 18 attributes. The first 4 attributes correspond to the date and time the meteorological data were measured. The remaining 14 attributes corresponds to wind direction, wind speed, gust speed, significant wave height, dominant wave period, average wave period, direction of dominant wave, sea-level pressure, air temperature, sea surface temperature, dewpoint temperature, station visibility, pressure tendency and water level. It is important to note that the wind speed is measured at a height of $3.8 - 5m$ above sea level.

### B. Tidal Current Data

The dataset used for calculating the output power of TEC is obtained from Center for Operational Oceanographic Products and Services [31]. The data is owned and maintained by National Oceanic and Atmospheric Administration. The datasets consist of measurements taken at 6-minute intervals over the course of several months for different locations across USA. Each row in the dataset corresponds to the measurements taken at a given time interval of the day. The dataset has 3 attributes. The first attribute for the date and time when the measurement was taken; the second attribute is the tidal velocity and the third column corresponds to the direction of the tide.

### C. Solar Irradiation Data

The dataset used for calculating the output power of the FPV system is obtained from the NREL PVWatts Calculator [32]. The dataset contains he hourly solar radiation and AC power after the inverter that can be generated from a PV panel in the given location.

## V. Projection of Offshore Renewable Generation

The governing equations to calculate the power output of each renewable energy resource are presented below. In this work, 3 regions were considered: Gulf of Mexico, Alaska and California.

### A. Output Power of TEC

For this work, a floating tidal turbine is considered. The TEC is rated at 500kW. Equation (14) represents the governing equation of electrical power output that can be obtained from WEC:

$$P_{TEC} = \frac{1}{2}\rho_t \pi r_t^2 (v_t)^3 \times C_p \times \eta \tag{14}$$

where $\rho_t$ is density of the fluid - mass of fluid per unit volume; $r_t$ is radius of the rotor; $v_t$ is tidal velocity; $C_p$ coefficient of power; and $\eta$ is efficiency of electrical system.

The hourly averages of the electrical power output obtained by a single TEC over a year for the 3 regions at each hour are shown in Fig. 1.

### B. Output Power of WEC

For this work, a linear attenuating WEC is considered. The WEC is rated at 750kW. The power matrix [33] is used to find the power output that a single linear TEC. The hourly averages of the electrical power output obtained by a single WEC over a year for the 3 regions at each hour are shown in Fig 2.

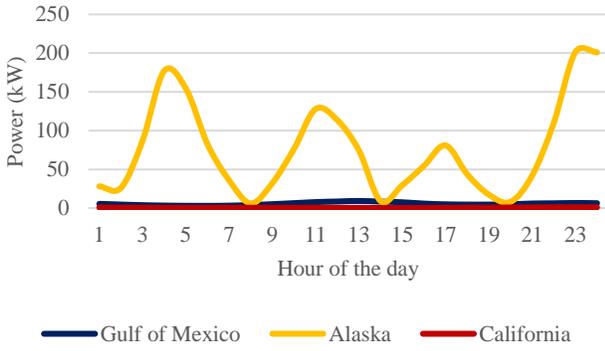

Fig. 1. Average power output of a single TEC of 500 kW.

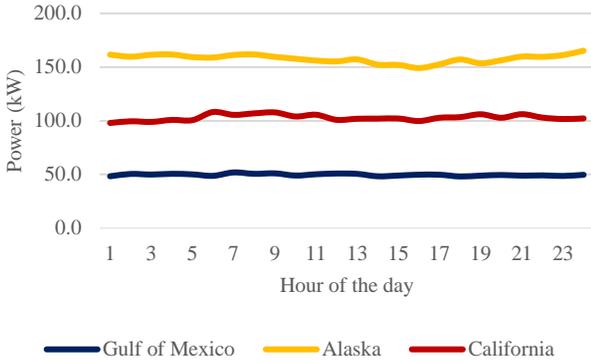

Fig. 2. Average power output of a single WEC unit of 750 kW.

### C. Output Power of OWT

As mentioned earlier in Section IV, the wind speed is measured at a height of 3.8-5m above sea level. However, the hub of OWT could be at a height of at least 80m above sea level. Thus, the log wind profile method [34]-[35] is used to estimate the wind speed at 80m height.

Equation (15) represents the governing equation of electrical power output that can be obtained from OWT. The size of a single OWT is rated at 8000kW:

$$P_{OWT} = \frac{1}{2}\rho_w \pi r_w^2 (v_w)^3 \times C_p \times \eta \qquad (15)$$

where, $\rho_w$ denotes density of air; $r_w$ is radius of the rotor; $v_w$ is the wind velocity; $C_p$ is the coefficient of power; and $\eta$ denotes the efficiency of electrical system.

The hourly averages of the electrical power output obtained by a single OWT over a year for the 3 regions at each hour are shown in Fig. 3.

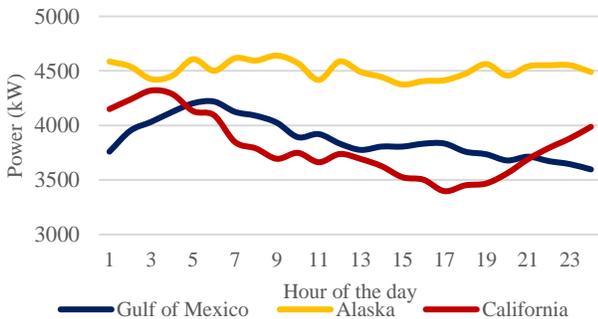

Fig. 3. Average power output of a single OWT unit of 8,000 kW.

### D. Output Power of FPV

The output power of the FPV was calculated using the NREL PVWatts Calculator. The size of a single FPV panel is set to 0.4kW. The hourly averages of the electrical power output obtained by a single FPV over a year for the 3 regions at each hour are shown in Fig. 4.

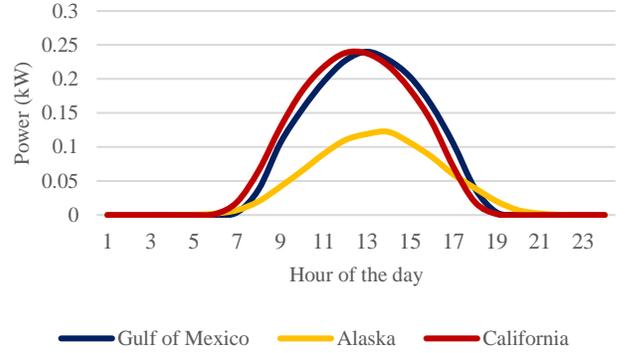

Fig. 4. Average power output of a single FPV of 0.4 kW.

## VI. Case Studies

An offshore oil project consists of four platforms, including one central power platform, one oil extraction and process platform, one oil storage platform and maintenance platform. The energy demand of O&G rigs could vary substantially. For instance, the peak power demand of the facilities in a rig could be around 44MW with a heating demand of more than 12 MW [36]. However, depending on the distance from shore and the number of heavy equipment employed in the rig, the electric power requirement can go up to 250 MW [37].

To demonstrate the proposed sizing model CORS for OHRES, a typical offshore platform is used as a test bed. The load profile of the test bed platform, shown in Fig. 5, does not fluctuate like residential loads since it operates continuously for 24 hours. The mining and oil processing activities constitute 80% of the load [38].

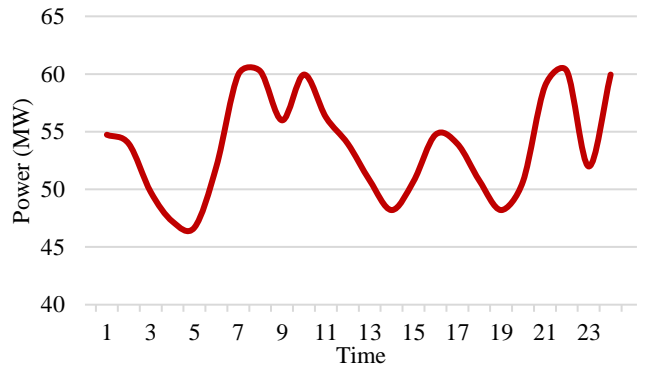

Fig. 5. Load profile for the offshore platform.

It is assumed in this paper that a typical day repeats itself for the next two decades. The case study is conducted on 3 different regions: Gulf of Mexico, Alaska, and California.

The proposed CORS model is implemented on Python using packages "Pyomo" [39] and "Gurobi" [40] was employed as the optimizer solver. The charging efficiency of the BESS is set at 80% and the discharging efficiency is set at 95%. The preset cost parameters are detailed in Table I.

Table I Cost parameters

| System | WEC (750kW) | TEC (500kW) | OWT (8000kW) | FPV (0.4kW) | BESS (kWh) |
|---|---|---|---|---|---|
| $C^{precom}$ ($) | 126,000 | 126,000 | 367,200 | 132 | 310 |
| $C^{capital}$ ($) | 6,300,000 | 6,598,500 | 16,038,767 | 520 | 150 |
| $C^{O\&M}$ ($ per annum) | 272,000 | 259,047 | 259,047 | 18 | 10 |
| $C^{decom}$ ($) | 1,000,000 | 0 | 1,123,333 | 35 | 100 |

Table II shows the results for the optimized lifetime cost of the proposed OHRES system in the 3 different regions. The cost was calculated for the total electricity consumption of 8,760,000MWh by an offshore platform over 2 decades.

Table II Optimized lifetime cost and quantity of energy resource units

| Regions | Gulf of Mexico | Alaska | California |
|---|---|---|---|
| Total Cost (M$) | 365.3 | 302.6 | 358.6 |
| WEC (nos.) | 0 | 0 | 0 |
| TEC (nos.) | 0 | 0 | 0 |
| OWT (nos.) | 14 | 13 | 15 |
| FPV (nos.) | 1,818 | 0 | 0 |
| BESS (kWh) | 49,923 | 8,136 | 19,758 |

In analyzing the OHRES in the Gulf of Mexico, Alaska and California, it is evident that the utilization of WEC and TEC was not deemed cost effective across any of these locations. Despite exhibiting a relatively stable output, the high associated costs prevented the consideration of WEC technology in all cases.

Among the three regions, Alaska stood out with its consistent power generation from OWT. With 13 turbines in place, Alaska experienced a steady and reliable output. Consequently, due to the dependable power supply from the OWT, the size of the BESS installed in Alaska could be much smaller.

## VII. CONCLUSION

A resource sizing model CORS for a OHRES system that could replace traditional electricity generation for offshore platforms with 100% clean energy and reduce $CO_2$ emissions is investigated.. Although the average electricity cost of the proposed system is currently higher than traditional diesel generators, the elimination of $CO_2$ emissions from offshore platforms makes it a viable alternative. Based on the proposed CORS model, the OHRES system is a feasible and reliable replacement for traditional systems, and its zero emission benefits contribute significantly to global decarbonization.

## VIII. FUTURE WORK

The ocean is a vast resource for renewable energy that can be harnessed for generating electricity. However, technological limitations, logistical, infrastructural and regulatory hurdles limit us from exploiting these resources to its fullest potential. Further research and development can help design the most efficient, yet cost effective and reliable offshore hybrid renewable energy systems.

For this work, it is assumed that the power requirement and generation of a single day would repeat itself for 20 years, which could be improved by considering the power generation and consumption requirements with higher resolutions. The OHRES system and the proposed CORS model considered 4 specific energy converters, however, they can be optimized to choose the best sizes of single candidate units of energy converters. Also, the model presented in this research only deals with 5 subsystems, more subsystems like hydrogen energy storage systems can be incorporated.

## ACKNOWLEDGMENT

This research is supported by Texas Commission on Environmental Quality through an award to the Subsea Systems Institute. This project was paid for [in part] with federal funding from the Department of the Treasury through the State of Texas under the Resources and Ecosystems Sustainability, Tourist Opportunities, and Revived Economies of the Gulf Coast States Act of 2012 (RESTORE Act). The content, statements, findings, opinions, conclusions, and recommendations are those of the author(s) and do not necessarily reflect the views of the State of Texas or the Treasury.